\documentclass{emulateapj}
\usepackage{graphicx}
\usepackage{amsmath}
\usepackage{amssymb}
\usepackage{epstopdf}
\usepackage{natbib}
\usepackage{color}
\usepackage{hhline}
\usepackage{subfigure}

\def\UW{\altaffilmark{1}}
\def\DUSC{\altaffilmark{2}}
\def\Penn{\altaffilmark{3}}
\def\INFN{\altaffilmark{4}}
\def\INAF{\altaffilmark{5}}
\def\OSU{\altaffilmark{6}}

\begin{document}

\title{Adding context to JWST surveys with current and future 21cm radio observations}
\shortauthors{Beardsley et al.}
\shorttitle{Adding context to JWST}

\author{A.~P.~Beardsley\UW$^,$\DUSC}
\author{M.~F.~Morales\UW$^,$\DUSC} 
\author{A.~Lidz\Penn}
\author{M.~Malloy\Penn}
\author{P.~M.~Sutter\INFN$^,$\INAF$^,$\OSU}

\altaffiltext{1}{Department of Physics, University of Washington, Seattle, WA 98195 USA}
\altaffiltext{2}{The Dark Universe Science Center, Seattle, WA 98195 USA}
\altaffiltext{3}{Department of Physics and Astronomy, University of Pennsylvania, Philadelphia, PA 19104, USA}
\altaffiltext{4}{INFN - National Institute for Nuclear Physics, via Valerio 2, I-34127, Trieste, Italy}
\altaffiltext{5}{INAF - Osservatorio Astronomico di Trieste, via Tiepolo 11, I-34143, Trieste, Italy}
\altaffiltext{6}{Center for Cosmology and Astro-Particle Physics, Ohio State University, Columbus, OH 43210, USA}

\email{beards@phys.washington.edu}

\begin{abstract}
Infrared and radio observations of the Epoch of Reionization promise to revolutionize our understanding of the cosmic dawn, and major efforts with the JWST, MWA and HERA are underway. While measurements of the ionizing sources with infrared telescopes and the effect of these sources on the intergalactic medium with radio telescopes \emph{should} be complementary, to date the wildly disparate angular resolutions and survey speeds have made connecting proposed observations difficult. In this paper we develop a method to bridge the gap between radio and infrared studies. While the radio images may not have the sensitivity and resolution to identify individual bubbles with high fidelity, by leveraging knowledge of the measured power spectrum we are able to separate regions that are likely ionized from largely neutral, providing context for the JWST observations of galaxy counts and properties in each. By providing the ionization context for infrared galaxy observations, this method can significantly enhance the science returns of JWST and other infrared observations.
\end{abstract}

\maketitle

\section{Introduction}
Although much of the evolution and history of the Universe is contained in the Cosmic Dark Ages and the Epoch of Reionization (EoR), these time periods remain largely unexamined. Current constraints of the EoR rely on indirect observations such as measurements of the polarization of the CMB, which provide only integral constraints on the reionization history \citep{Komatsu:2011}, and spectroscopy of highly redshifted quasars \citep{Mortlock:2011}. Deep galaxy surveys in the optical and infrared are beginning to constrain the end of reionization with a few dozen high redshift star forming galaxies \citep{Bouwens:2010}.

Several ground-based radio experiments designed to probe the EoR through power spectrum measurements have recently come online, including the GMRT (Giant Metrewave Radio Telescope, \citealt{Paciga:2013}), LOFAR (LOw Frequency Array\footnote{http://www.lofar.org}, \citealt{Yatawatta:2013}), PAPER (Precision Array for Probing the Epoch of Reionization\footnote{http://eor.berkeley.edu}, \citealt{Parsons:2010}), and the MWA (Murchison Widefield Array\footnote{http://www.mwatelescope.org}, \citealt{Tingay:2012}). These instruments aim to exploit the hyper-fine transition of the neutral hydrogen residing in the intergalactic medium (IGM) to study the reionization history. Due to an extremely faint signal and high thermal noise, the first generation of these experiments was designed to perform a statistical measurement of the EoR in the form of the cosmological power spectrum. The second generation of such experiments is just on the horizon with the Hydrogen Epoch of Reionization Array (HERA\footnote{http://reionization.org}), which is also designed to perform a power spectrum measurement but with much higher sensitivity and design decisions informed by first generation experiments (DeBoer, et al. 2014, in prep). While the first generation of radio arrays will not have sufficient sensitivity to produce a full image of the EoR, recent studies of instrumental effects and sensitivity have motivated the possibility of imaging at very large scales \citep{Malloy:2013, chapman:2013}. 

Meanwhile, the JWST (James Webb Space Telescope\footnote{http://www.jwst.nasa.gov/}) is in development to explore the EoR in the infrared regime with a much higher resolution and smaller field of view compared to the ground based radio projects \citep{gardner:2006}. This instrument will have sufficient sensitivity and resolution to identify and study individual galaxies during the EoR (e.g. \citealt{Zackrisson:2011}). Due to its small field of view ($\sim2$ arcmin), the JWST cannot perform a comprehensive survey of the EoR. Instead it will need to focus on small patches of the sky to collect a representative sample of the Universe's evolution. A 21cm map of the IGM will then allow JWST to correlate the ionization fraction of the gas with properties such as luminosity functions, spectral energy distributions, morphologies, and the emission strengths of lines like Ly-$\alpha$ or H$\alpha$. These correlations in turn can answer questions about how the environment of galaxies affect how they form and evolve.

Below we demonstrate the potential to produce such a map using a current radio array (MWA) and a future array (HERA) to provide information about the IGM environment of the galaxies which will be studied by the JWST. We use a realistic HI signal simulation and take into account several dominant instrumental effects when creating our map. There is potential for similar cross-studies with other experiments such as the LSST deep drilling fields \citep{Lazio:2014}, a WFIRST Guest Observer program \citep{Spergel:2013}, or existing Hubble Space Telescope fields \citep{Giavalisco:2004}. \cite{Lidz:2009} explored the merit of a cross power spectrum between 21cm maps and an extended Subaru survey \citep{Kashikawa:2006}, however Subaru has very small overlapping survey area with selected fields for deep integration by the MWA, and no overlap with HERA. We focus on the JWST as one of its core science programs is to study the EoR, and it will produce very deep pointed images at the redshifts of interest.

The remainder of this paper is organized as follows: in Section \ref{sec:sim} we describe the simulation used, in Section \ref{sec:instrument} we describe our instrumental model including thermal noise and foreground contamination, in Section \ref{sec:results} we show the results of our imaging plan, and we conclude and discuss future work in Section \ref{sec:conclusion}. 

Throughout this paper we use a $\Lambda$CDM cosmology with $\Omega_m=0.73$, $\Omega_\Lambda=0.27$, and $h = 0.7$, consistent with WMAP seven year results \citep{Komatsu:2011}. All distances and wavenumbers are in comoving coordinates.

\section{The 21 cm Signal}\label{sec:sim}
To show the capability of the MWA and HERA to image the EoR we use a simulation of the 21 cm signal. We make use of a ``semi-numeric" reionization simulation (e.g. \citealt{Zahn:2007, Mesinger:2011}) based on excursion set modeling of the EoR \citep{Furlanetto:2004}. Specifically, we use the models from \cite{Malloy:2013}, which track the 21cm brightness temperature field across 512$^3$ cells in a periodic, 1 co-moving Gpc/h simulation box.

The contrast between 21 cm brightness temperature and the CMB at a given location, $\mathbf{r}$, can be approximately expressed in terms of the fractional baryon over density, $\delta_\rho$, and the local ionization fraction, $x_i$, as
\begin{equation}
\delta T_b(\mathbf{r}) = T_0(1-x_{i}(\mathbf{r}))(1+\delta_\rho(\mathbf{r}))
\end{equation}
where $T_0=28[(1+z)/10]^{1/2}$ mK. We will use a simulation cube of $\delta T_b$ as the signal to be detected by our instrument, and the corresponding cube of $x_i$ to inform us of the underlying ionization environment. We also use this simulation to make a power spectrum to use in our filter (Eq. \ref{eq:filter}). The simulation box is 1 $h^{-1}$Gpc on a side, while the MWA field of view at $z_{\text{fid}}=6.9$ is much larger. To account for this we tile the simulation cubes. The periodic boundary conditions ensure no artificial discontinuities are introduced.

For this study we use a mean ionization fraction $\bar{x}_i=0.79$ as a benchmark because the bubbles are sufficiently large to do some imaging, but there is still enough neutral gas around to leave some contrast in the brightness temperature field. A slice of constant $r_{||}$ (line of sight) of the brightness temperature cube is shown in Figure \ref{fig:input_image}. The characteristic bubble size in this image is significantly smaller than the expected resolution of the MWA at this sensitivity. However, we aim to extract meaningful information about ionization environment without explicitly identifying bubbles or characterizing their sizes.

\begin{figure}
\begin{center}
\includegraphics[width=\columnwidth]{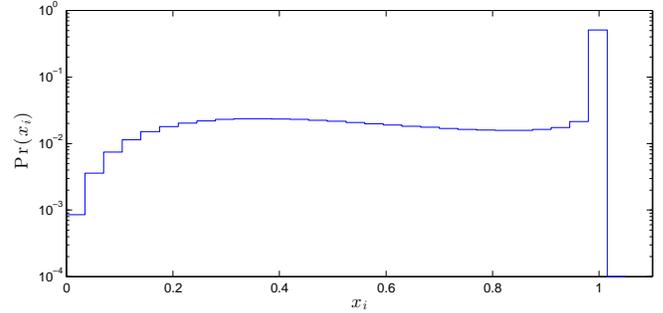}
\end{center}
\caption{Histogram of ionization fraction $x_i$ for the fiducial input simulation. More than half (51\%) of the pixels are fully ionized ($x_i>0.98$), while the remaining pixels have a very broad and flat distribution ranging from fully neutral to nearly ionized.}
\label{fig:xi_hist}
\end{figure}

As a reference for upcoming plots, a histogram of the ionization fraction per pixel of the fiducial input simulation is shown in Figure \ref{fig:xi_hist}. By a happy coincidence, at the redshifts studied here, the simulation pixel size is approximately the same size as the field of view of JWST (1.1 arcmin and 2.2 arcmin respectively).  The per pixel ionization fraction should be interpreted as a smearing to the scale of the simulation pixel size, or half of a JWST field. With sufficient resolution (scale of individual galaxies), every pixel would either be fully ionized or fully neutral, and the distribution would simply have 79\% pixels fully ionized. Because of our finite sized pixels, we instead have a two distributions. Approximately half of pixels reside in bubbles larger than the pixel size and are fully ionized ($>98\%$, last bin), while the majority of the remaining pixels have a very broad distribution spanning fully neutral ($x_i=0$) to nearly completely ionized ($x_i \lesssim 0.98$). 

A blind JWST survey would probe galaxies that reside in IGM with the population shown in Figure \ref{fig:xi_hist}. While any random pointing without any prior information has about a 51\% chance of being completely ionized and 23\% chance of being mostly neutral, the galaxy surveys will not be able to differentiate between IGM environments. Our goal is to distinguish the regions of fully ionized gas from those less than half ionized even without fully resolving the reionization bubble distribution, providing two separate populations for the JWST to correlate against.

\section{Instrument model}\label{sec:instrument}
We use a realistic model of the current MWA taking into account several instrumental effects to model the thermal noise and the foreground contamination. We outline the highlights of this model here, and a detailed description can be found in \cite{Beardsley:2013}. The observing parameters used can be found in Table \ref{tbl:obs_params}.

\begin{table}
\begin{center}
\caption{Observational parameters\label{tbl:obs_params}}
\begin{tabular}{lcc}
\hline\hline
Parameter & MWA & HERA \\
\hline
No. of antennas & 112* & 37, 127, 331\\
Central frequency & 180 MHz (z $\sim$ 6.9) & 180 MHz\\
Field of view & 27.2$^{\circ}$ & 7.3$^{\circ}$\\
Effective area per antenna & 20 m$^2$ & 154 m$^2$\\
Total bandwidth & 6 MHz & 6 MHz\\
T$_{\mathrm{sys}}$ & 315.5 K & 315.5 K\\
Channel width & 80 kHz & 96 kHz\\
Latitude & -26.701$^{\circ}$ & -30.0$^{\circ}$\\
Integration per day & 6 hr & 32 min\\
Total integration & 1,000 hr & 120 hr\\
\hline
\multicolumn{3}{p{\columnwidth}}{*Sixteen of the 128 MWA antennas lie significantly far from the core to provide high resolution. However, in the very low signal to noise regime these antennas offer very little sensitivity and greatly increase the size of the $uv$ plane, so are omitted from this work.}
\end{tabular}
\end{center}
\end{table}

Each visibility measurement of an interferometer is sensitive to a Fourier mode perpendicular to the line of sight, given by $\mathbf{k}_\perp=2\pi\mathbf{u}/D_m(z)$ where $\mathbf{u}$ is the baseline vector in units of wavelengths and $D_m(z)$ is the transverse comoving distance to the observation at redshift $z$ \citep{Hogg:1999}. Additionally, the frequency dimension maps to $r_{||}$ through the cosmological redshift due to the expansion of the Universe. The thermal noise on each visibility is
\begin{equation}\label{eq:thermal_noise}
V_{\text{rms}}(\mathbf{k}_{\perp},r_{||})=\frac{c^2 T_{\text{sys}}}{f^2 A_e \sqrt{\Delta f \tau}}
\end{equation}
where $T_{\text{sys}}$ is the system temperature, $A_e$ is the effective collecting area per antenna, $\Delta f$ is the frequency channel width, and $\tau$ is the integration time. This can easily be expanded to a long tracked observation by allowing $\tau$ to represent the total time observing a given $(\mathbf{k}_{\perp},r_{||})$ bin including redundant baselines and accounting for rotation of the earth.

We can then create a noisy image by adding Gaussian random noise at the level of Eq. \ref{eq:thermal_noise} to the Fourier transform of the simulation brightness temperature cube, where $\tau$ is independently calculated for each pixel.
\begin{equation}\label{eq:noisy_data}
\delta\tilde{T'_b}(\mathbf{k}) = \delta\tilde{T_b}(\mathbf{k})+\tilde{n}(\mathbf{k})
\end{equation}
Here we use a prime to represent a measurement of a signal and a tilde to represent a variable in Fourier space. In the step from Eq. \ref{eq:thermal_noise} to Eq. \ref{eq:noisy_data}, we implicitly performed a Fourier transform of the noise along the line of sight. This is a straightforward operation with uniform noise in the direction of the transform (see \cite{Morales:2005} for a full derivation).

Next we apply a filter to the measurement which will do two things: it will mask out modes which are expected to be contaminated by foregrounds, and it will down-weight low signal to noise modes.

\begin{figure}
\begin{center}
\subfigure[]{\includegraphics[width=\columnwidth]{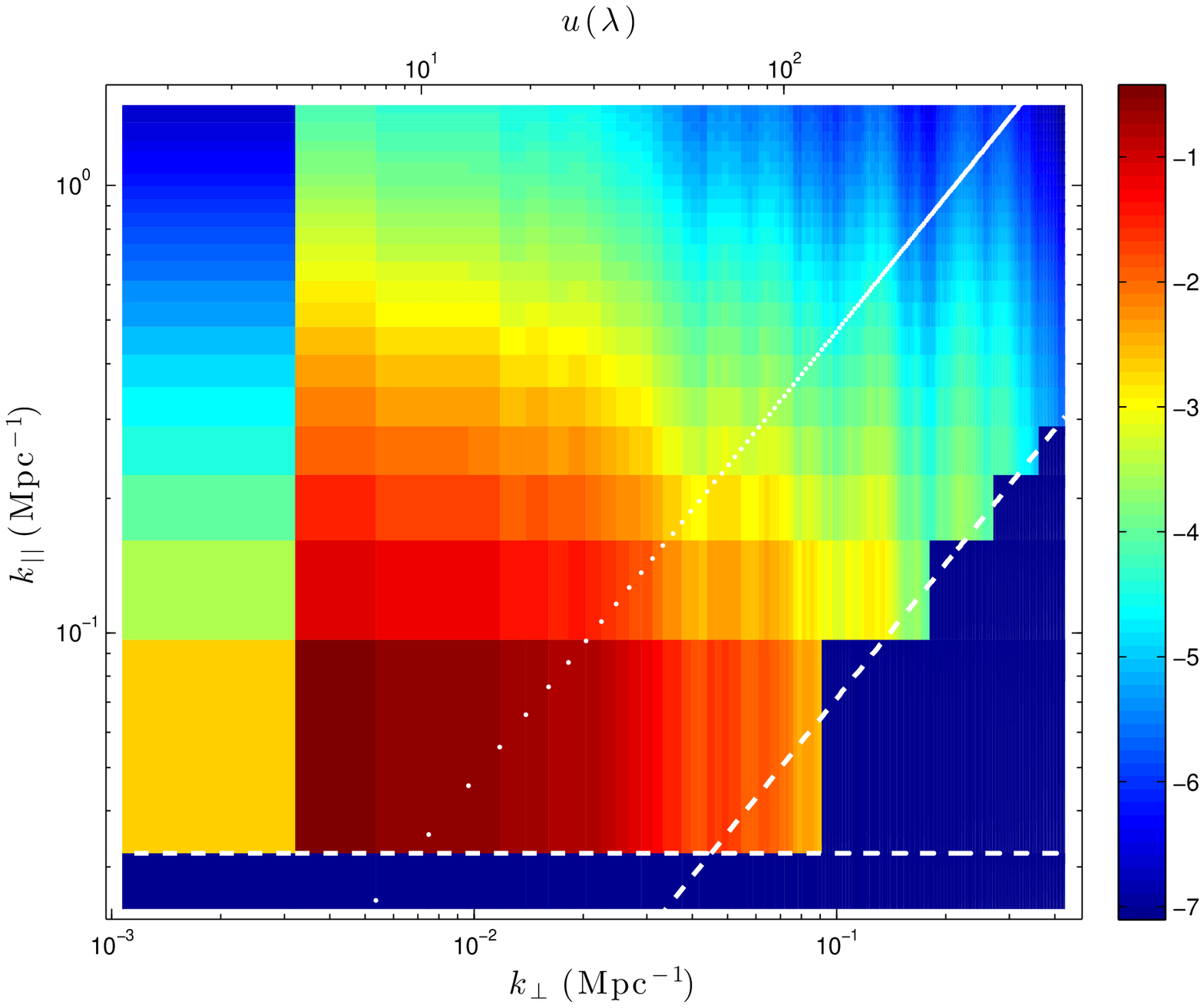}\label{fig:filter_k}}
\subfigure[]{\includegraphics[width=\columnwidth]{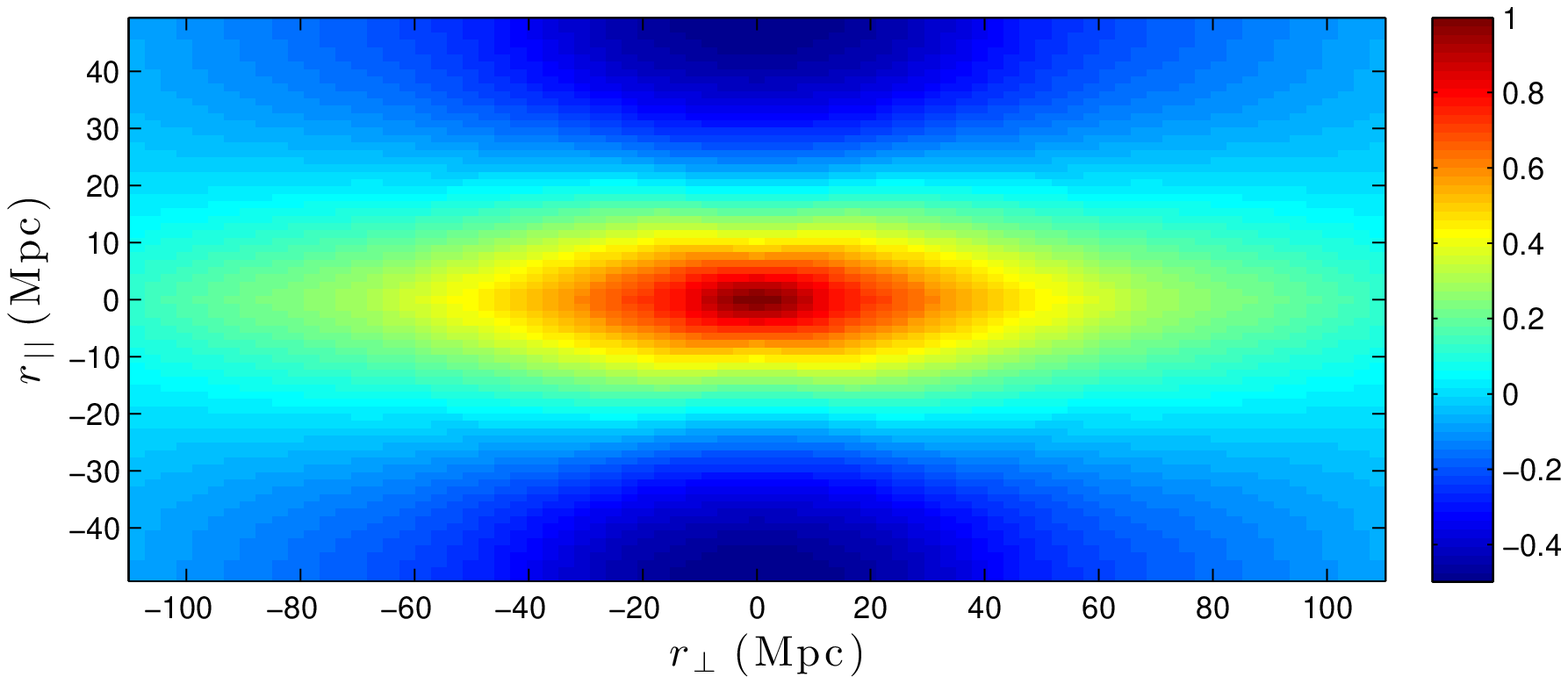}\label{fig:filter_r}}
\end{center}
\caption{The filter used for our fiducial test. \emph{Top:} A slice of the filter, $\log_{10}(F(k_\perp,k_{||})$), in wavenumber space. A log color scale was used to emphasize the bins contributing to the filter. The bins below the horizontal white dashed line and to the right of the diagonal white dashed line are masked out due to foreground contamination, and shown in dark blue. Most of the sensitivity of our measurement lies at low $k_\perp$, low $k_{||}$. The sharp drop in sensitivity at the lowest $k_\perp$ bin is due to the minimum spacing of antennas in the MWA. \emph{Bottom}: The image space representation of the filter, $F(r_\perp,r_{||})$, or the inverse Fourier transform of the top image. Shown is a slice of constant $r_x=0$. Along the line of sight direction the filter ranges from negative response to a positive peak back to negative. The response is elongated in the perpendicular direction due to the smaller $k_\perp$ modes contributing.}
\label{fig:filter}
\end{figure}

Foregrounds are expected to be a major challenge for radio observations of the EoR, however the effects of foreground contamination have been studied in great detail and shown to be restricted to low $k_{||}$ modes and a so-called ``wedge" (see, e.g. \citealt{Liu:2014a,Liu:2014b, pober:2013,hazelton:2013,thyagarajan:2013,trott:2012,morales:2012,vedantham:2012,datta:2010,bowman:2009, morales:2006}). How far the wedge bleeds into high $k_{||}$ and whether cosmological information can be extracted is an active topic of investigation. We adopt a fairly optimistic fiducial and use the field of view as the limit of foreground contamination, but explore other choices in the following section. The foreground mask is shown in Figure \ref{fig:filter_k}.

\begin{figure*}
\begin{center}
\subfigure[]{\includegraphics[width=1.0\columnwidth]{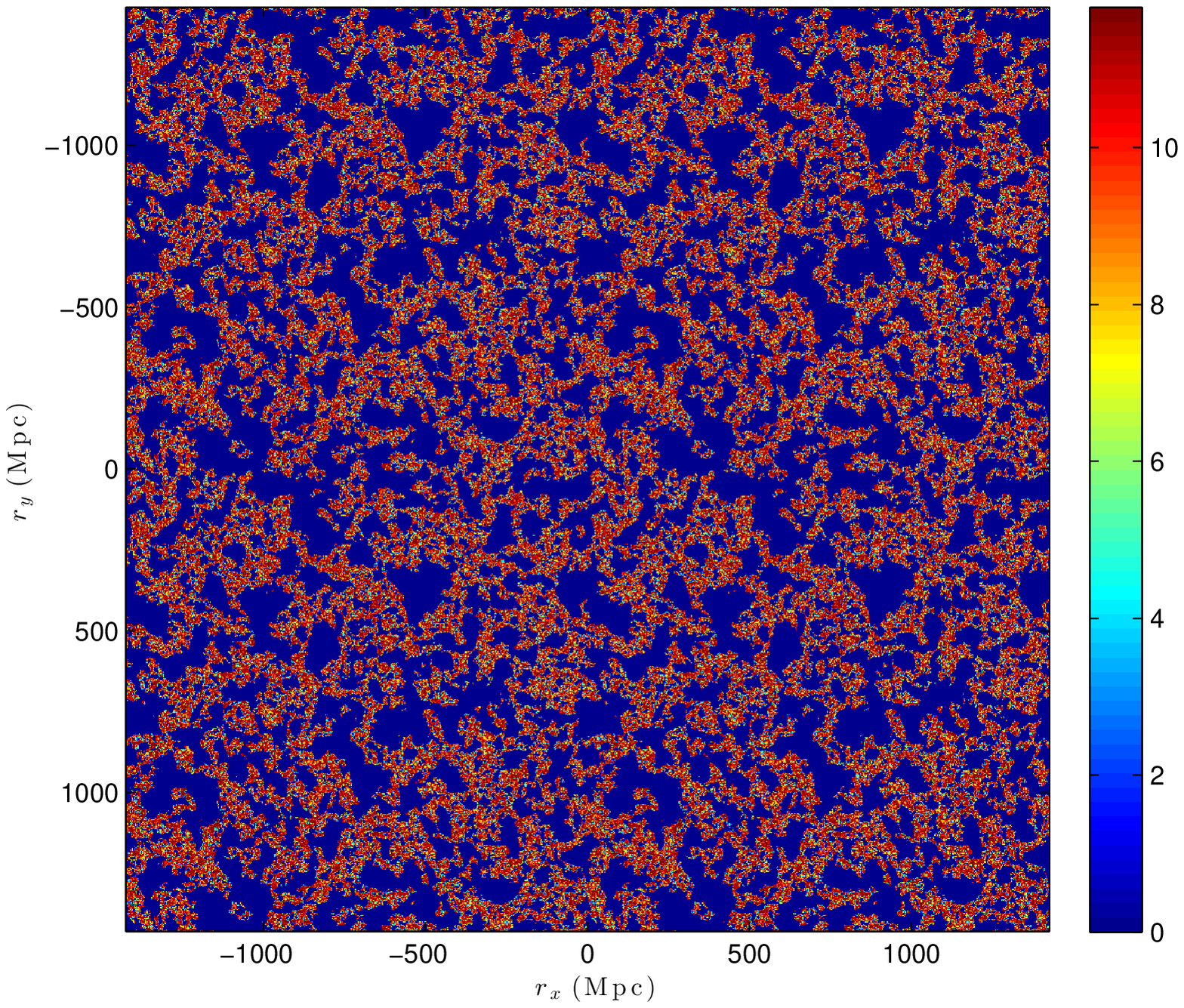}\label{fig:input_image}}
\subfigure[]{\includegraphics[width=1.0\columnwidth]{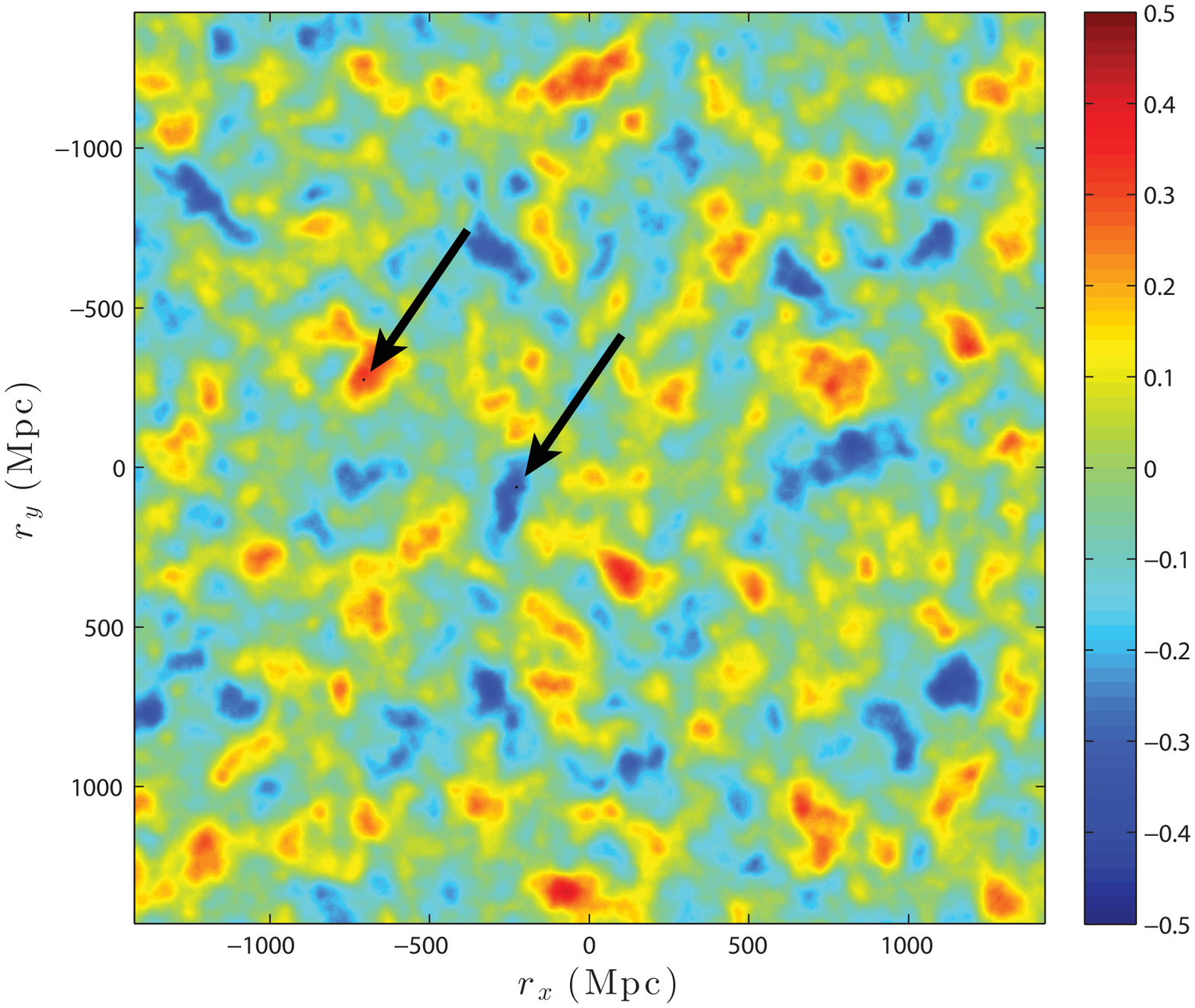}\label{fig:filtered_image}}
\end{center}
\caption{\emph{Left:} The input simulation brightness temperature, $\delta T_b(\mathbf{r})$, in mK. The dark blue regions are zero temperature and represent fully ionized bubbles. \emph{Right:} The filtered noisy image produced by our fiducial MWA instrument model, in arbitrary units. Owing to the peculiar shape of the filter (Figure \ref{fig:filter}), the structure of the filtered image can look very different than the input image. The arrows are pointing to tiny black boxes approximately the size of the JWST field of view. A full infrared survey would probe many such regions of varying IGM ionization fractions, informed by the probability distributions shown in Figure \ref{fig:prob_xi_wedge}.}
\label{fig:images}
\end{figure*}

To down weight low signal to noise modes we use the Wiener filter, which is defined by the power spectra of the signal, $P_{21}(\mathbf{k})$, and the noise, $P_{N}(\mathbf{k})$. The signal power spectrum will be measured from actual observation data, but here is estimated from the simulation cube itself. It is important to note here that the form of our filter, and thus subsequent analysis, is dependent on the HI power spectrum. For our fiducial test, the signal power spectrum is calculated from the simulation image shown in Figure \ref{fig:input_image}. Figure 1 of \citealt{Lidz:2008} explores the form and evolution of the HI power spectrum. While the methodologies, assumptions, and simulation size used there are different than our current simulation, the power spectra are qualitatively similar and the differences have minimal impact on our results. We explore the dependence of our results on the ionization fraction (which in turn can be used as a proxy for bubble sizes) in Section \ref{sec:results}. The noise power spectrum comes from propagating Eq. \ref{eq:thermal_noise} to power spectrum space. The resulting filter can be expressed as
\begin{equation}\label{eq:filter}
F(\mathbf{k})=
\begin{cases}
0 & \text{if contaminated}\\
P_{21}(\mathbf{k})/(P_{21}(\mathbf{k})+P_N(\mathbf{k})) & \text{otherwise.}
\end{cases}
\end{equation}

Our estimate of the EoR image can then be written as the inverse Fourier transform of the filter multiplied by our $k$ space measurement.
\begin{equation}\label{eq:image_estimate}
\delta\hat{T}_b(\mathbf{r})=N\times{\cal FT}^{-1}\left[F(\mathbf{k})\delta\tilde{T'_b}(\mathbf{k})\right]
\end{equation}
We also introduce a normalization factor, $N$, to set the scale of the resulting image. Because interferometers produce zero-mean images, and much of the power has been removed due to our filter, we normalize each filtered noisy image to have mean of zero and range of one with arbitrary units. This forces values to be between roughly -0.5 and 0.5, and allows us to plot images with very different filters on the same axes (see Section \ref{sec:results}).

The shape of this filter in $k$ space is shown in Figure \ref{fig:filter_k}. The horizontal dashed white line represents the low $k_{||}$ contaminated by foregrounds. Because the MWA observes with an instantaneous bandwidth of $30.72$ MHz but the cosmological measurement is restricted to 6 MHz, the $\sim4$ modes of expected contamination are restricted to only the lowest $k_{||}$ bin in our case. The right-most diagonal dashed white line represents the wedge characterized by the field of view of the MWA. All bins below the horizontal and to the right of the diagonal dashed line are masked out, and shown as dark blue.

The majority of the sensitivity lies at low $k_\perp$ and low $k_{||}$. This is due to a combination of lower thermal noise due to a dense array core and the signal power spectrum being high for small $k$. The sharp drop off of sensitivity at low $k_\perp$ is due to not having antennas placed physically closer than 7.5 m apart. However, there is \emph{some} sensitivity there due to projection from earth rotation.

An image space representation of our filter is shown in Figure \ref{fig:filter_r}. The slice shown is for constant $r_x$=0 to show the shape in $r_{||}$ and $r_\perp$. In this space the field of view is much larger than the line of sight extent, so the image has been cropped in $r_\perp$ but not $r_{||}$. Because the majority of line of sight sensitivity comes from the first non-zero $k_{||}$ mode, the line of sight response resembles a single mode curve moving from negative response at the near edge, peaking in the center, and returning to negative at the far edge. In addition the response in the perpendicular direction is much more broad than the line of sight. 

An example of a filtered noisy image is shown in Figure \ref{fig:filtered_image}. Due to the peculiar filter shape and natural sensitivity of the instrument, the images contain very non-intuitive structures. Indeed it can be extremely difficult to gauge the quality of the images without a rigorous metric. In the following section we show that despite low signal to noise and a highly irregular response, we can retrieve information about the underlying ionization field.

\begin{figure}
\begin{center}
\includegraphics[width=\columnwidth]{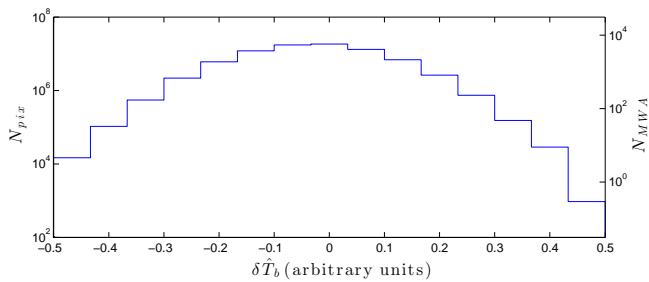}
\end{center}
\caption{Histogram of the values in the filtered noisy image shown in Figure \ref{fig:filtered_image}, using our fiducial MWA instrument model. The total number of pixels in the simulation is $\sim8\times10^7$. The second vertical axis provides a reference to the approximate number of independent regions in the map, given the filter size for the MWA.}
\label{fig:filtered_image_hist}
\end{figure}

For reference we show a histogram of the measured temperature values in Figure \ref{fig:filtered_image_hist}. For each temperature value, $N_{\text{pix}}$ is the number of simulation pixels measured at that value. The pixels in our filtered map are not all independent owing to the relatively large point spread function of our filter, so we provide a second vertical axis ($N_{\text{MWA}}$) showing the approximate number of independent regions in the map, determined by the size of the MWA filter.

\section{Results}\label{sec:results}
In this section we attempt to answer a simple question: Given a measured value in a noisy filtered 21 cm image, what, if anything, can we say about the ionization fraction of the IGM for the galaxies that JWST would observe at that location?

\begin{figure}
\begin{center}
\includegraphics[width=\columnwidth]{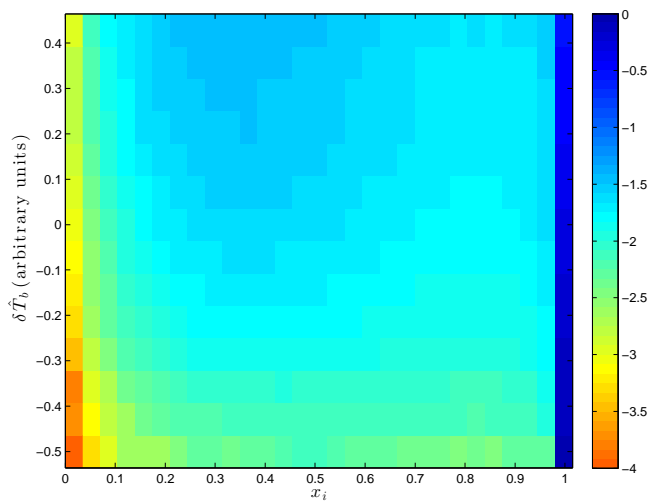}
\end{center}
\caption{Image correlations with $x_i$ (log color scale). For each binned observed temperature, $\delta\hat T_b$, we histogram the underlying pixels in the simulation $x_i$ cube and normalize each row independently to a sum of one. Each row of the image can be viewed as the probability distribution of ionization fraction given a observed temperature. Most of the distribution lies at high $x_i$ due to the highly peaked native distribution (Figure \ref{fig:xi_hist}). Nevertheless, a correlation can be seen between measured value and input ionization fraction.}
\label{fig:corr_mat}
\end{figure}

To answer this question, we first bin the values of the filtered noisy image into fifteen equal sized bins. Then for each binned temperature value, we construct the probability distribution of the underlying ionization fraction by histogramming the $x_i$ values corresponding to the selected image pixels. The result for our fiducial MWA observation are shown in Figure \ref{fig:corr_mat}.

Due to the extremely peaked input distribution (Figure \ref{fig:xi_hist}), the resulting correlations are also heavily skewed toward high $x_i$. In other words, no matter what value is measured in our image, there is a significant probability it is a fully ionized region simply because most of the image is fully ionized, and our fiducial instrument does not have the resolution to decouple small pockets of ionized gas from neutral clouds. 

But not all is lost. Two encouraging features can be seen in Figure \ref{fig:corr_mat}. The first is that the highly ionized bin on the far right drops by about a half order of magnitude (note the log color scale - the corresponding values are 0.8 to 0.2) moving up in the figure. This means that a low observed temperature is much more likely to be fully ionized than a high value. The second feature is the gradual increase in probability of low ionization moving to higher brightness temperatures. However, like the input image, the distribution is quite broad and not peaked in any one place. In an attempt to quantify these features, we further bin the ionization fractions. We wish to ignore the broad, mildly ionized pixels and focus on two distinct sub-populations of pixels: those that are fully ionized ($x_i>0.98$) and those that are less than half ionized ($x_i<0.5$). These populations account for 51\% and 23\% of the input simulation pixels, respectively. We will find that division leads to a clearer distinction in the underlying $x_i$ probability distributions.

\begin{figure}
\begin{center}
\includegraphics[width=\columnwidth]{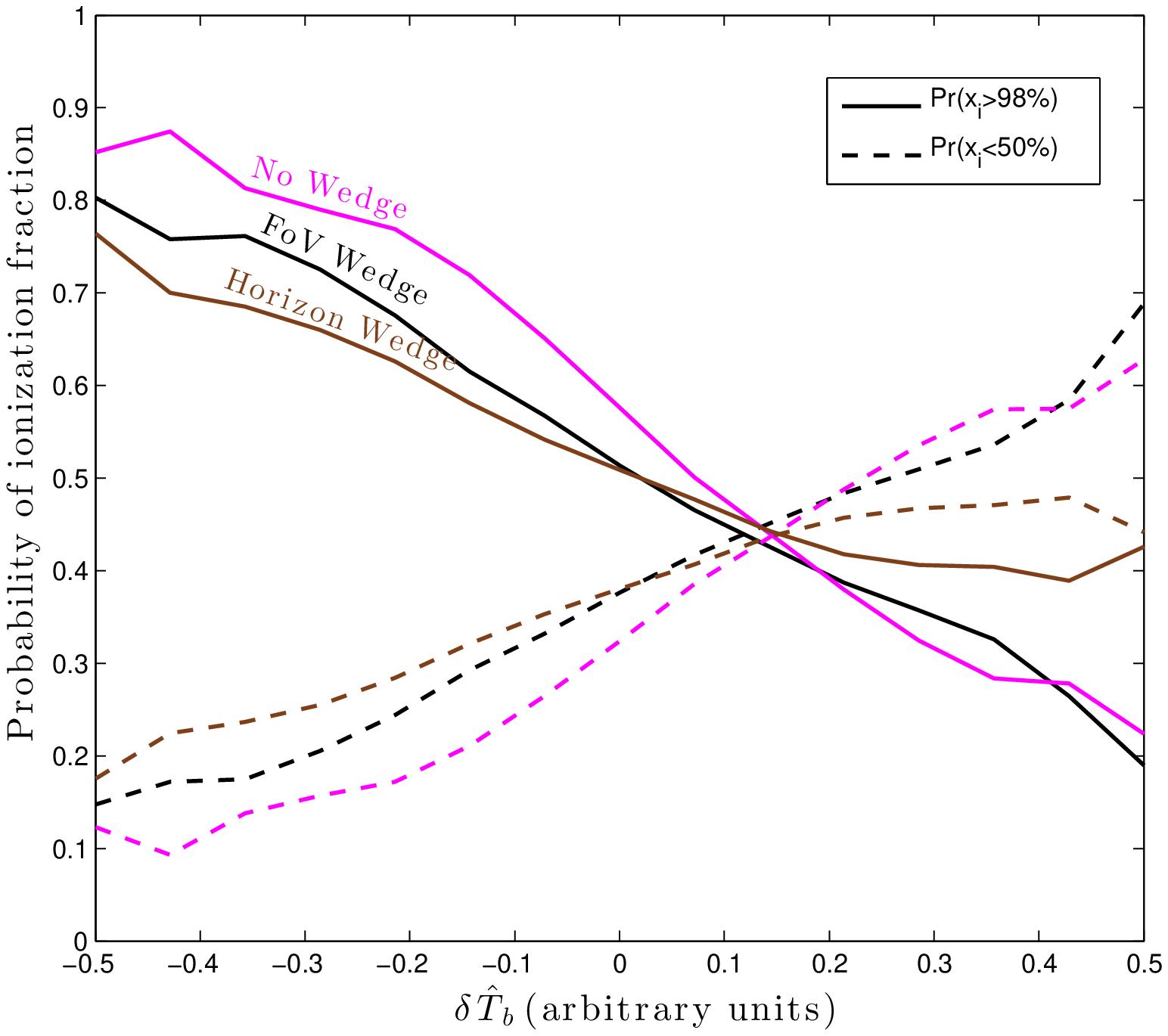}
\end{center}
\caption{The probability of fully ionized pixels (solid lines) and mostly neutral pixels (dashed lines) as a function of measured temperature, $\delta\hat{T}_b$. The separation, cross, and re-separation of the solid and dashed lines show that our method is able to distinguish between the two populations of ionized environments. The black lines represent our fiducial MWA instrument with mean ionization fraction of 0.79 and a foreground wedge defined by the field of view. Also plotted are the very optimistic no wedge (magenta lines) and the pessimistic horizon wedge (brown lines).}
\label{fig:prob_xi_wedge}
\end{figure}

In Figures \ref{fig:prob_xi_wedge}-\ref{fig:prob_xi_hera} the solid line of a given color is the probability distribution of fully ionized pixels, while the dashed line of the same color corresponds to less than half ionized pixels. The colors then represent different trials which we explain in turn.

First let us examine the fiducial observation indicated by the black lines in Figure \ref{fig:prob_xi_wedge} and reproduced in the subsequent figures for reference. This trial was done using the MWA observation parameters listed in Table \ref{tbl:obs_params}, a wedge defined by the field of view, and a mean ionization fraction of 79\%. The most striking feature is the large separation in probabilities at low observed temperature, with the probability of fully ionized far exceeding the ``blind" guess of 51\%. Also encouraging to note is that moving up in filtered temperature, the solid and dashed black lines cross and separate again. This means that we can potentially distinguish two regions of ionization levels with our filtered noisy image. 

We will see in the following sub-sections that the success of our filter to distinguish ionization regions is closely tied to the size of the point spread function compared to the size of the bubbles in the simulation. When the filtered instrument has sensitivity on order of the size of ionized bubbles, it tends to have success in identifying ionized regions which have low temperature due to the lack of neutral hydrogen. Similarly, where the instrument can resolve large regions of neutral hydrogen, we have success on the right side of the plot. It is not surprising that we tend to have higher certainty for ionized regions than we do for neutral regions because ionized bubbles tend to saturate with extended regions of purely ionized gas, while neutral clouds contain an admixture of neutral and partially ionized regions that current radio experiments cannot hope to resolve.

While we cannot claim with certainty the ionization fraction of any given pixel, we can produce a partial sky map of ionized and non-ionized regions based on a given probability threshold. With such a map, the JWST can form an observing plan to probe regions of high and low IGM ionization fraction, allowing the study of galactic properties in different environments. In Figure \ref{fig:images} we show schematically a couple fields that JWST could observe to probe a highly ionized region and a mostly neutral region. A comprehensive survey would involve many such fields, statistically incorporating our probability distribution functions.

\subsection{The Influence of the Wedge}
Many factors can influence the effectiveness of this strategy. Here we explore a few and leave a comprehensive study to future work. The first variable we explore is a choice of the wedge. Our fiducial choice is fairly optimistic and relies on no foreground contamination outside the field of view of the instrument. 

Figure \ref{fig:prob_xi_wedge} shows the probability lines for two other choices as well. First is the extremely optimistic choice of there being no foregrounds at all. While there is no realistic expectation that current or future radio observations will be able to completely remove the wedge and recover all cosmological information from those modes, we include this line to show the effect of going from no foregrounds whatsoever to including a more realistic wedge. Indeed the effect is quite minimal. While the no wedge line shows a slightly higher probability of full ionization at low measured values, it has a crossing at roughly the same location as the field of view wedge line, and has very similar values on the right side of the plot. 

Although quite unexpected, this surprisingly small effect can be understood by examining the filter shown in Figure \ref{fig:filter_k}. While there are many bins excluded below the white dashed lines, the strong majority of our sensitivity actually lies safely outside the wedge at low $k_\perp$. So while adding information will almost always help, it is easy to see why in this case the difference between a field of view wedge and no wedge at all is marginal.

Next we consider a pessimistic wedge defined by the horizon. The dotted white diagonal line in Figure \ref{fig:filter_k} represents the cut made in Fourier space for this choice. Anything below that line is assumed to be contaminated by foregrounds and omitted. Here we can see the wedge digs into the $k_\perp$ sensitivity of the instrument, and so we expect a poorer resolution. In Figure \ref{fig:prob_xi_wedge} we see the effect on the correlation with $x_i$ values. The separation at low observed temperatures is marginally worse than the field of view wedge case. However, the separation at high temperature values is essentially nonexistent. The poor resolution due to foreground contamination has washed out any ability of our instrument to distinguish small pockets of neutral gas. 

Even with a pessimistic outlook on foreground mitigation, a separation of distributions at low $\delta\hat{T}_b$ offers valuable information for a JWST survey. But to get the most out of these images, it is necessary to continue investigation into foreground subtraction and avoidance to try to recover as much of the wedge as possible.

\subsection{Mean Ionization Fraction}
Next we turn to uncertainty in the ionization history. Although the power spectrum measurement will be able to constrain models of reionization, it is not clear how well the mean ionization fraction $\bar{x}_i$ for a given redshift will be known (see Pober, et al. 2014, in prep., for a recent discussion). We explore this uncertainty by producing images for simulation cubes at mean ionization fractions of $\bar{x}_i=0.68$ and 0.89 in addition to our fiducial value of 0.79.

\begin{figure}
\begin{center}
\includegraphics[width=\columnwidth]{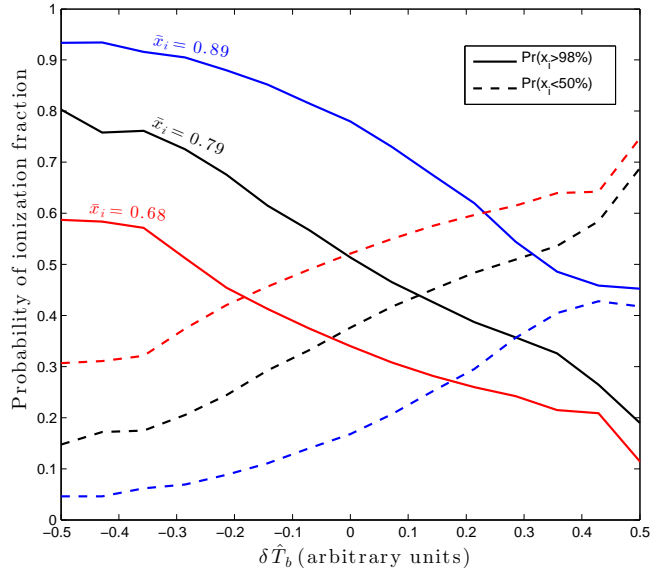}
\end{center}
\caption{Probabilities of ionization populations for varying mean ionization fraction. The fiducial lines are reproduced in black, while the blue lines show a more ionized Universe ($\bar{x}_i=0.89$) and the red lines represent the less ionized $\bar{x}_i=0.68$. As in Figure \ref{fig:prob_xi_wedge}, the solid lines represent fully ionized regions while the dashed lines represent less than half ionized.}
\label{fig:prob_xi}
\end{figure}

The results are shown in Figure \ref{fig:prob_xi}. The effect of a higher than expected ionization fraction is a wider separation at low $\delta\hat{T}$, but no cross over between populations. This can be explained with the fact that the input image is dominated by large ionized bubbles. Our instrument has low resolution, so is better at picking out large bubbles (low observed temperatures), but the remaining neutral clouds are limited to small isolated regions to which our instrument is blind.

On the other hand, a lower mean ionization yields a small separation at low $\delta\hat{T}_b$ and large separation at large values. The explanation for this effect is the same as the latter but run in reverse. At low ionization fraction the bubbles are smaller and our instrument has a harder time detecting them. But the neutral clouds are large and easily distinguished.

From this test we learn the importance of considering the uncertainty of ionization fraction. While any one of these lines can provide valuable IGM information to JWST surveys, a full marginalization over uncertainty should be performed to ensure accurate statistics. The details however, are outside the scope of this paper and ultimately depend on the detected power spectrum and the uncertainty it yields on ionization fraction. To rough approximation, if the ionization fraction can be known to $\pm0.1$, the red and blue curves can be interpreted as the error bars on the black curve.

\subsection{Future Observations with HERA}
Finally, we set our sights on the next generation of radio EoR experiments. HERA was designed to have much larger collecting area per element as well as highly redundant baselines for increased sensitivity. The larger area per element results in a more narrow field of view. Additionally, the elements cannot point and thus HERA is naturally a drift instrument resulting in far less integration time per day on a single field. The dishes are arranged in a closely packed hexagonal configuration. Current preliminary funding for HERA is sufficient to build 37 dishes in South Africa, while additional funding could expand to 331 dishes. The observing parameters for our HERA model are shown in the right column of Table \ref{tbl:obs_params}. Note the drastically smaller integration time used for HERA compared to the fiducial MWA.

\begin{figure}
\begin{center}
\includegraphics[width=\columnwidth]{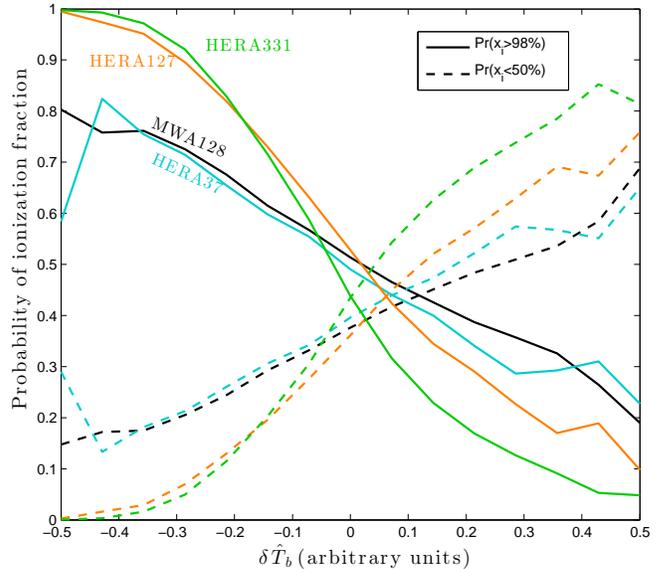}
\end{center}
\caption{Probabilities of ionization populations for the HERA instrument. The currently funded 37 dish HERA (cyan lines) perform comparable to the fiducial MWA, while further buildouts of 127 dishes (orange lines) and 331 dishes (green lines) strongly outperform our fiducial and offer the most information for correlations with JWST.}
\label{fig:prob_xi_hera}
\end{figure}

The results of our HERA simulation are shown in Figure \ref{fig:prob_xi_hera}. With only 120 hours of integration, the currently funded 37 dish instrument performs very comparable to the fiducial MWA with 1,000 hours. Adding more dishes substantially improves the sensitivity. With an intermediate 127 dishes, HERA will be able to identify ionized regions with near certainty, and an improved ability to distinguish neutral regions (high values of $\delta\hat{T}_b$). This may seem counter-intuitive due to the shorter baselines of the tightly-packed HERA compared to the MWA. But due to the large elements and redundant configuration, HERA actually has high sensitivity out to larger baselines than the MWA, resulting in a better resolution after the filter is applied. Therefore it is more capable of distinguishing ionized from neutral regions.

Finally, as expected the fully proposed 331 dish HERA will perform the best of our tests. Not only can it identify fully ionized regions with near 100\% certainty like the 127 dish version, but it can also identify mostly neutral regions to $\gtrsim$80\% confidence. The increased confidence on either end of the distribution will help constrain the correlations with JWST galaxy surveys.

Our HERA calculations have been performed on a single field of view region of the sky. Because HERA is a drift instrument, it will observe many such fields every day resulting in a strip of sky at this level of sensitivity.

\section{Conclusions and Further Work}\label{sec:conclusion}
We have shown that current radio interferometers, despite being designed for power spectrum measurements, can produce useful image information that can provide context to future galaxy surveys by instruments like the JWST. The probability distribution functions we have shown in Figures \ref{fig:prob_xi_wedge}-\ref{fig:prob_xi_hera} will allow for a statistical study of local ionization fraction against galaxy properties such as luminosity functions, spectral energy distributions, and morphologies probed by infrared surveys. Such studies will enable us to answer questions such as:
\newpage\begin{itemize}
\item How does the apparent age of stars in galaxies depend on ionization environment?
\item What is the dust content of galaxies residing in neutral environments versus ionized?
\item How did reionization affect the shapes and sizes of galaxies forming from ionized gas?
\end{itemize}

In our analysis the foreground contaminated ``wedge" proved to be a less significant factor than expected. While there exist gains between the horizon wedge and the field of view wedge, the difference between field of view and no wedge at all is only marginal. The highly active investigations of foreground contamination will hopefully allow imaging to push down to the field of view.

Our technique is also robust against uncertainties in mean ionization fraction. A more comprehensive marginalization over $\bar{x}_i$ is needed before correlations with future surveys can be performed. However, the marginalization will depend strongly on the constraints provided by power spectrum measurements and other experiments.

Our method may be improved by considering other filters and imaging strategies. The Wiener filter was used here as a benchmark to yield high signal to noise, however further optimization may be achieved by considering the probability distributions of Figures \ref{fig:prob_xi_wedge}-\ref{fig:prob_xi_hera} as a figure of merit. Here we avoided attempts to make claims on the sizes of the ionized bubbles and instead focused on an instrument-driven filter, but additional work with matched filters akin to \cite{Malloy:2013} will yield additional scientifically interesting information.

An additional followup to this work would be to correlate the filtered noisy images with other galaxy properties in the simulations. For example interesting cosmological and astrophysical information is contained in the actual timing of reionized bubbles ($z_{re}$) or the number of galaxies with halos over some threshold mass within ionized regions. Measurements of this variety would give more handles on the underlying reionization models.

The MWA has completed its first year of EoR observing, and has recently begun its second year which will nearly triple the total observing time. The work currently being done to eliminate systematics and reduce the data to a power spectrum measurement is essential in the path towards our proposed imaging project. Meanwhile HERA is pushing forward and will begin construction in 2015. This next generation instrument will build on the lessons learned thus far and provide a much more sensitive power spectrum measurement, as well as potential for imaging. Observations from these instruments will allow us to provide IGM context to the deep galaxy surveys of instruments like the JWST which will study the objects reionizing the Universe.

\section*{Acknowledgments}
The authors would like to thank Matthew McQuinn, Bryna Hazelton, and Jonathan Pober for extremely useful discussions that led to this work. MFM and APB acknowledge support through NSF grants AST-0847753, AST-1410484, and AST-1206552. AL and MM acknowledge support through NSF grant AST-1109156. PMS is supported by the INFN IS PD51 ``Indark".

\bibliography{../Thesis/beardsley}
\bibliographystyle{apj}

\end{document}